\documentclass[9pt,twocolumn,twoside]{pnas-new_arxiv}
\templatetype{pnasresearcharticle}

\title{Universality of local weak interactions and its application for interferometric alignment}

\author[a,b]{Jan Dziewior}
\author[a,b]{Lukas Knips}
\author[c]{Demitry Farfurnik\textsuperscript{1}}
\author[a,b]{Katharina Senkalla}
\author[c]{Nimrod Benshalom}
\author[c]{Jonathan Efroni}
\author[a,b]{Jasmin Meinecke}
\author[c]{Shimshon Bar-Ad}
\author[a,b]{Harald Weinfurter}
\author[c]{Lev Vaidman\textsuperscript{2}}
\affil[a]{Max-Planck-Institut f\"{u}r Quantenoptik, Hans-Kopfermann-Stra{\ss}e 1, 85748 Garching, Germany}
\affil[b]{Department f\"{u}r Physik, Ludwig-Maximilians-Universit\"{a}t, 80797 M\"{u}nchen, Germany}
\affil[c]{Raymond and Beverly Sackler School of Physics and Astronomy, Tel-Aviv University, Tel-Aviv 69978, Israel}
\leadauthor{Dziewior}

\significancestatement{
In this combined theoretical and experimental work we describe and demonstrate a universal feature of local couplings of spatially pre- and postselected systems.
All weak couplings, independently of their physical nature, are modified in the same universal manner characterized by the weak value of local projection.
We present a general analysis which neither relies on a specific form of the coupling nor is restricted to weak interactions.
This allows to define a new consistent paradigm of the presence of a quantum particle, much more intricate than the concept of the presence of a classical particle.
Furthermore, we discovered that Gaussian pointers remain reliable beyond the weak regime, which enabled a novel, simple, and efficient alignment method for interferometric setups.
}

\authorcontributions{J.D., L.K., H.W., and L.V. designed research; J.D., L.K., K.S., and L.V. performed research; J.D., L.K., D.F., and L.V. contributed new analytic tools; J.D., L.K., D.F., K.S., N.B., J.E., S.B.-A., and L.V. analyzed data; J.D., L.K., K.S., N.B., J.M., H.W., and L.V. wrote the paper; D.F., N.B., J.E., and S.B.-A. performed preliminary experimental work on the alignment method; D.F., N.B., J.E., and S.B.-A. participated in the analyses; and D.F. suggested the alignment method.}
\authordeclaration{The authors declare no conflict of interest.}
\equalauthors{\textsuperscript{1}Current Address: Racah Institute of Physics, The Hebrew University of Jerusalem, Jerusalem 9190401, Israel; The Center
for Nanoscience and Nanotechnology, The Hebrew University of Jerusalem, Jerusalem 9190401, Israel}
\correspondingauthor{\textsuperscript{2}To whom correspondence should be addressed. E-mail: vaidman@post.tau.ac.il}


\begin{abstract}
The modification of the effect of interactions of a particle as a function of its pre- and postselected states is analyzed theoretically and experimentally.
The universality property of this modification in the case of local interactions of a spatially pre- and postselected particle has been found.
It allowed to define an operational approach for the characterization of the presence of a quantum  particle in a particular place: the way it modifies the effect of local interactions.
The experiment demonstrating this universality property provides an efficient interferometric alignment method, in which the position of the beam on a single detector throughout one phase scan yields all misalignment parameters.
\end{abstract}

\begin{document}

\maketitle
\thispagestyle{firststyle}
\ifthenelse{\boolean{shortarticle}}{\ifthenelse{\boolean{singlecolumn}}{\abscontentformatted}{\abscontent}}{}

\section{Introduction}
Pre- and postselected systems are ubiquitous in quantum mechanics.
In many quantum information schemes the intended process is only realized by the interplay of preselection and postselection.
The addition of postselection, often together with conditioned transformations, is the basis of protocols such as universal quantum computation within the Knill-Laflamme-Milburn scheme~\cite{Knill2001}, entanglement swapping~\cite{Zukowski1993} and heralding in general~\cite{Zeilinger1997}.

The two-state vector formalism (TSVF)~\cite{AVreview} provides a general framework for the description of pre- and postselected systems.
It introduces a state evolving backwards in time and thereby treats the postselection on equal footing as the pre\-selection.
The key element of the TSVF is the weak value of an observable.
As long as the interaction is sufficiently weak or short the observable effect on the external system is completely characterized by the weak value~\cite{beyond}.
For such interactions, the state of the external systems after the postselection can deviate significantly from the states expected by just considering the coupling to preselected systems~\cite{AAV}.
The concept of weak values became the basis of several successful applications in precision measurement techniques~\cite{Hosten2008,Dixon2009}.
While there are theoretical controversies about the optimality of the weak value-based tomography and precision-measurement methods~\cite{Zilberberg2011,Wu2012,Hofmann2012a,Xu2013,Dressel2014,Jordan2014,Ferrie2014,Knee2014,Magana-Loaiza2014,Pusey2014,Zhang2015a,Piacentini2018} a plethora of fruitful applications continues to emerge~\cite{Martinez-Rincon2017a,Li2017,Araujo2017,Qiu2017,Liu2017,Chen2018,Kim2018,Zhou2018,Li2018,Qin2018,Ren2018,Huang2018,Fang2018,Li2018a}.

We take a step back and investigate the fundamental properties of pre- and postselected systems.
We find that there exists a general universality principle characterizing how the effects of the interactions in one location of a spatially pre- and post-selected quantum system are modified as a function of pre- and postselection.
All these modifications are specified by a single complex number, the weak value of the spatial projection operator.
One of the innovations of our approach is that it does not rely on the specific form of the interaction Hamiltonian.
Instead, it expresses the change of the state via the complex amplitude of an orthogonal component, which emerges due to the interaction.
If the weak value is a positive number, the size of the changes in every variable is multiplied by this number and when it is negative, all modifications happen in the opposite direction.
If the effect originally changed a particular variable, in the case of an imaginary weak value, the effect will occur in a variable conjugate to the initial one, and when the weak value is a complex number, both effects are combined together.
This approach allows a formal definition of a quantum particle's presence.

Until now, most accounts considered the weak value to be limited to the case of weak interactions, e.g.~\cite{Wu2011,DiLorenzo2012,Kofman2012,Zhang2016,Denkmayr2017}.
It is another crucial innovation of our approach, however, that we explicitly apply the formalism to the case of much stronger interactions.
We use an expression for the weak value which takes into account changes due to interactions of finite strength in the time interval between pre- and postselection.
Besides incorporating the stronger interactions we also account for decoherence or imperfections in the measurement system.
We show experimentally that this weak value can in fact be measured using weakly coupled pointers.

An interferometer, especially a Mach-Zehnder type interferometer, can be seen as the iconic example for pre- and postselected systems.
The reflectivity/transmittivity of the first beamsplitter together with the phase shifter defines the preselected state of the system.
The final beamsplitter together with detection of the particle in one output of the interferometer sets the postselected state.
The effect of weak interactions of the particle with external systems, which can be seen as a trace the particle leaves inside the interferometer, is characterized by the weak value of the projection operator on the corresponding arm.
Surprisingly, we also found that for Gaussian states of the external system, the weak value characterizes the modification of the trace for arbitrary strength of the interaction.

The interferometer enables straightforward experimental implementation where we consider a pre- and postselected photon passing through.
We experimentally characterize the various effects of multiple interactions in one of the interferometer's arms using the mode and the polarization of the propagating photon as the external systems coupled to the photon's path.
We find that the modifications of the weak effects on the photon can be described by the weak value of the projection operator on the corresponding arm for various types and strengths of couplings.

We can now turn the picture upside down and view any coupling to the external degrees of freedom as being due to misalignment of the interferometer.
For example, a tilted mirror in one of the beams now becomes an interaction deflecting the Gaussian mode of the beam from its ideal direction.
This analogy directly leads us to an efficient alignment technique for interferometers where our analysis provides a simple model for the image observable at the output of an interferometer.
More precisely, by measuring the phase dependent trajectory of the centroid of the output mode on only a single spatially resolving detector we can extract the misalignment parameters in one go.
This technique harnesses the benefits of the weak amplification method~\cite{AAV} to improve precision.

\section{Weak value of local projection and its connection to the trace} \label{sec::wv_ideal}
Let us first consider the effect of a quantum particle on external systems due to all kinds of local interactions in the channel through which it passes.
The interactions might be caused by various properties of the particle, e.g., charge, mass,  magnetic moment, etc., but we assume that the particle passing through the channel does not change its quantum state.\footnote{To deal with cases, where the states of some degrees of freedom of the particle change, we treat them equivalently to the external degrees of freedom.
In fact, this is the case in our experiment, see Sec.~\ref{sec::univ}.}

\begin{figure}[t]
\includegraphics[width=0.45\textwidth]{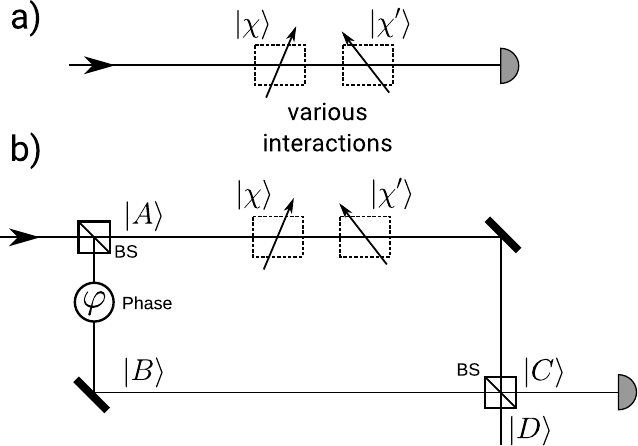}\\
\caption{\textbf{Comparison between effect of the local coupling of the particle when it passes through a single channel and when it passes through an interferometer.} a).
The particle interacts with external systems in a single channel originally in the state $|\chi\rangle$.
b) The particle passes through a Mach-Zehnder interferometer with arm $A$ identical to the channel described in (a) with all its local interactions,  while it is assumed that there are no local couplings in arm $B$.
}
\label{fig::MZI_basic}
\end{figure}

If the quantum particle is not present in the channel, the state of the external systems at a particular time is $| \chi \rangle$.
When the quantum particle is localized in the channel as shown in Fig.~\ref{fig::MZI_basic}a, the interactions change the total state of the external systems as
\begin{equation}\label{eq::iaSingle}
|\chi \rangle \rightarrow|\chi^\prime \rangle \equiv \eta\left( |\chi \rangle + \epsilon |\chi^\perp \rangle \right),
\end{equation}
where $| \chi^\perp \rangle$ denotes the component of $| \chi^\prime \rangle$ which is orthogonal to $| \chi \rangle$.
By definition we choose the phase of $|\chi^\perp\rangle$ such that $\epsilon > 0$.
For simplicity, but without loss of generality we also disregard the global phase and consider the coefficient $\eta$ to be positive such that $\eta = \langle \chi | \chi^\prime \rangle = \frac{1}{\sqrt{1+\epsilon^2}}$.
The trace left by the particle is manifested by the presence of the orthogonal component $| \chi^\perp \rangle$ and is quantified by the parameter $\epsilon$.

Next, let this channel be an arm of a Mach-Zehnder interferometer (MZI), see Fig.~\ref{fig::MZI_basic}b. We assume that the arm $B$ of the MZI is ideal, i.e., the particle leaves no trace there.\footnote{This assumption is made for simplicity of presentation. The main results of the paper about the universality of modification of interactions are easily transformed to the case when some weak traces are left in all parts of the interferometer.}
For the creation of the preselected state inside the interferometer $| \psi \rangle$ the unbalanced input beam splitter is followed by a phase shifter resulting in
\begin{align}\label{eq::preSelSimp}
| \psi \rangle = \cos \alpha | A \rangle +  \sin \alpha e^{i\varphi} | B \rangle,
\end{align}
where $| A \rangle$ and $| B \rangle$ represent the eigenstates of the path degree of freedom, and $\alpha$ and $\varphi$ are the two real parameters of the state.

The second beam splitter is balanced, so its operation can be modeled as
\begin{subequations}
\begin{align}
| A \rangle &\rightarrow \frac {1}{\sqrt 2} ( | C \rangle + | D \rangle), \\
| B \rangle &\rightarrow \frac{1}{\sqrt 2} ( | C \rangle - | D \rangle).
\end{align}
\end{subequations}
We collect photons in output port $C$, which corresponds to a postselection of the state
\begin{align}\label{eq::postSelSimp}
|\phi \rangle = \frac{1}{\sqrt{2}}\left( | A \rangle + | B \rangle \right).
\end{align}

Accounting for the interactions in arm $A$ (see Fig.~\ref{fig::MZI_basic}b) the composite state $| \Psi \rangle$ of the particle and the external systems before the second beam splitter is
\begin{align} \label{eq::stateFinal}
|\Psi \rangle  &=  \cos \alpha | A \rangle | \chi^\prime \rangle + \sin \alpha e^{i\varphi} | B \rangle | \chi \rangle,
\end{align}
where here and in the rest of the paper we employ a shorthand notation for tensor products with $| A \rangle | \chi^\prime \rangle \equiv | A \rangle \otimes | \chi^\prime \rangle$.
After detection of the particle in arm $C$, i.e., postselection of the particle in state [\ref{eq::postSelSimp}], the state of the external systems becomes
\begin{align}\label{eq::compPostSel}
|\tilde{\chi} \rangle =\mathcal{N} \left( |\chi \rangle + \frac{\eta \epsilon}{\eta + \tan \alpha e^{i\varphi}} |\chi^\perp \rangle \right),
\end{align}
where $\mathcal{N}$ is the normalization factor.
Here and in the rest of the paper we use the accent symbol ``$\sim$'' to denote situations with pre- and postselection.

We start by considering interactions which are sufficiently small, with $\epsilon \ll 1$.
In the case of a single channel, the particle passing through it leads to the change of the state of the external systems,
\begin{equation}\label{eq::iaAmp1}
| \chi \rangle \rightarrow |\chi^\prime  \rangle = |\chi \rangle +\epsilon  |\chi^\perp \rangle + \mathcal{O}(\epsilon^2),
\end{equation}
which is just an expansion of [\ref{eq::iaSingle}] in orders of $\epsilon$.

For the particle that has passed through the corresponding MZI and has been detected in $C$ we observe a different change of the state of the external systems.
Expanding Eq.~[\ref{eq::compPostSel}] in orders of $\epsilon$ we can see that the weak effect of the interaction is modified relative to [\ref{eq::iaAmp1}] by a single parameter, the weak value of projection on arm $A$,
\begin{equation}\label{eq::iaAmp}
| \chi \rangle \rightarrow |\tilde{\chi} \rangle=|\chi \rangle +\epsilon \left( \mathbf{P}_A  \right)_w |\chi^\perp \rangle + \mathcal{O}(\epsilon^2),
\end{equation}
where, for defining the weak value, we neglect the coupling to external systems
\begin{align}\label{eq::wvIdealMZI}
\left( {\rm \bf P}_A \right)_w &\equiv \frac{ \langle \phi | {\rm \bf P}_A | \psi \rangle }{\langle \phi | \psi \rangle} = \frac{1}{1 + \tan \alpha ~ e^{i\varphi}}.
\end{align}

The design of the interferometer allows the full range of weak values of projection onto arm $A$, by varying the parameters $\tan \alpha$ and $\varphi$.
Note that we did not restrict the number of interactions as long as their combined effect is sufficiently weak.

When the trace left in the interferometer is small, ${\epsilon \ll 1}$, the weak value can be considered neglecting the effect of the interactions as in [\ref{eq::wvIdealMZI}].
In the next Section we will turn towards scenarios with stronger couplings for which the interactions cannot be neglected.

\section{Weak Value considering finite coupling strength and imperfections}
\label{sec::wv_finite}
Calculating the weak value as in Eq.~[$\ref{eq::wvIdealMZI}$] we have implicitly assumed that it only depends on the pre- and postselection states at the boundaries of the considered time interval.
This is correct in the limit of weak coupling, which is considered in most works about weak measurements.
Yet, sometimes even in scenarios with coupling of finite strength the weak value has been treated as if there was no coupling, i.e., using formula [$\ref{eq::wvIdealMZI}$] \cite{Wu2011,DiLorenzo2012,Kofman2012,Zhang2016,Denkmayr2017,Piacentini2018,Vaidman2017a}.

To correctly account for couplings of finite strength, we turn to the proper definition of the weak value in the framework of the TSVF, which refers to a single point in time $t$, at which the particular forward and backward evolving quantum states have to be evaluated \cite{AV90}.
All interactions of finite strength and imperfections of optical devices between preselection and $t$ as well as between $t$ and postelection, must be considered.
Thus, Eq.~[\ref{eq::preSelSimp}] correctly describes the forward evolving state only immediately after the first beam splitter and Eq.~[\ref{eq::postSelSimp}] describes the backward evolving state only immediately before the second beam splitter.
Since all evolutions due to imperfections or interactions with the different external systems are local, i.e, they have the common eigenstates $|A\rangle$ and $|B\rangle$, the time ordering of the evolutions is of no consequence.
Therefore, the weak value $(\textbf{P}_A)_w$ stays constant in time and we are free to choose any moment in time to calculate it.
For convenience, we calculate the weak value immediately before postselection on state [\ref{eq::postSelSimp}] and modify only the forward evolving state to account for the evolution due to interactions inside the interferometer.

Due to the interactions the system becomes entangled with the external systems as described by Eq.~[\ref{eq::stateFinal}].
Thus, the particle is in the mixed state described by the density matrix in the basis $\left\lbrace | A \rangle, | B \rangle \right\rbrace$
\begin{align}\label{eq::preSelEff}
\rho = \begin{pmatrix}
\cos^2 \alpha & \cos \alpha \sin \alpha e^{-i\varphi} \eta \\
\cos \alpha \sin \alpha e^{i\varphi} \eta & \sin^2 \alpha
\end{pmatrix}.
\end{align}

The weak value in the case of mixed states has been derived in \cite{beyond} (Eq.~(32) therein),
\begin{align} \label{eq::mixedWV}
A_w = \frac{\mathrm{Tr} \left( \rho_\text{post} A \rho_\text{pre} \right)}{\mathrm{Tr} \left( \rho_\text{post} \rho_\text{pre} \right)}.
\end{align}
In our case this formula is not applicable for arbitrary time between the pre- and postselection due to entanglement in both forward and backward evolving states with the same external systems, see Section VI of \cite{beyond}, but it can be used to calculate the weak value immediately before the last beam splitter since the backward evolving state is not entangled, see also \cite{Wu2011,Wiseman2002,Silva2014}.
As we explained above, the weak value of the projector $\mathbf{P}_A$ is constant between pre- and postselection, so it can be calculated as
\begin{align}\label{eq::wvModified}
\left( \mathbf{P}_A  \right)_w = \frac{\mathrm{Tr} \left( |\phi \rangle  \langle \phi| \mathbf{P}_A \rho \right)}{\mathrm{Tr} \left( |\phi \rangle \langle \phi| \rho \right)}
= \frac{1 + \tan \alpha \, \eta e^{-i\varphi}}{1 + \tan^2 \alpha + 2 \tan \alpha \, \eta \cos\varphi}.
\end{align}

\begin{figure*}
\includegraphics[width=1\textwidth]{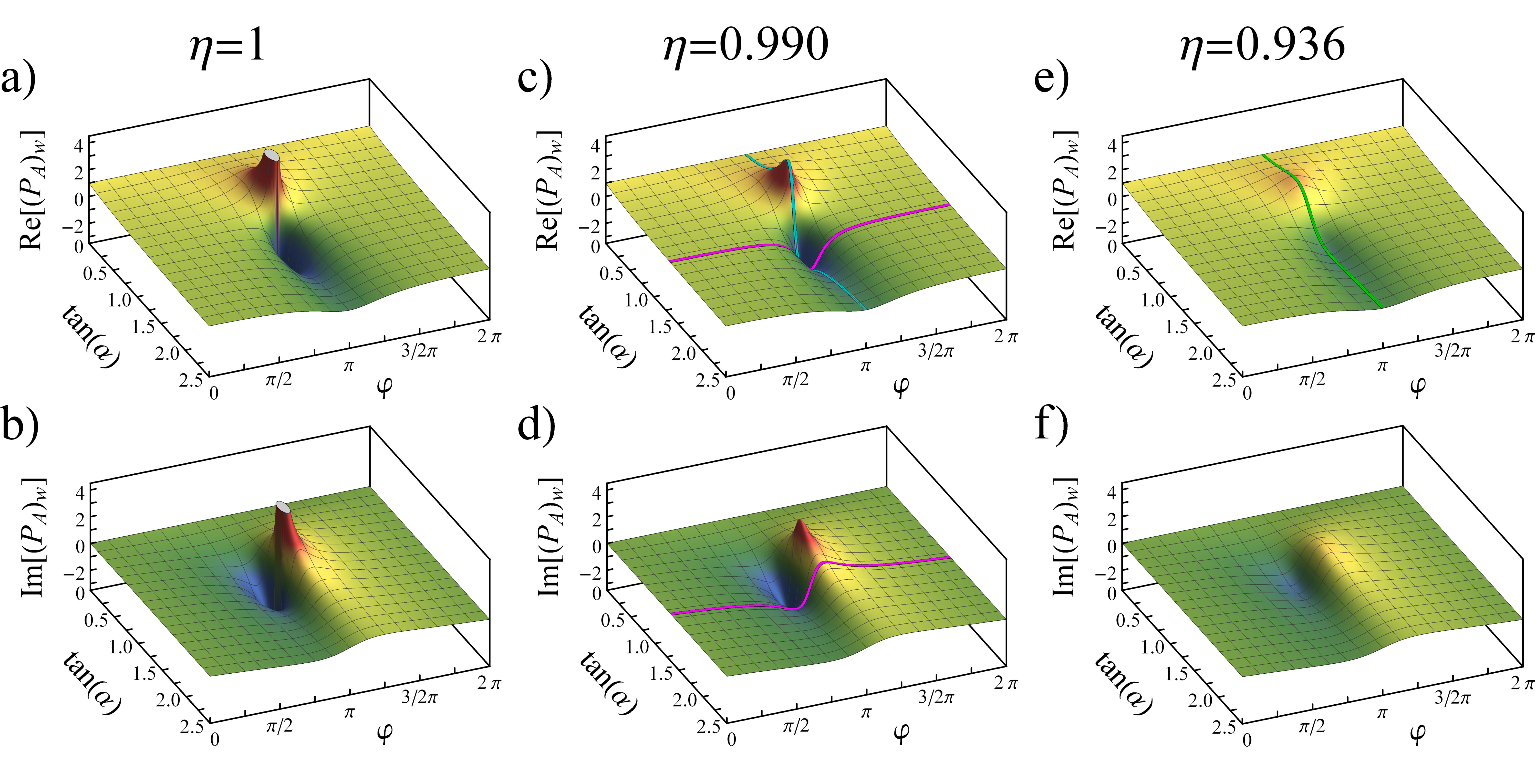}
\caption{\textbf{Exact parameter dependence of weak value.}
Real (upper row) and imaginary (lower row) parts of weak value of the projection operator on arm $A$ for $\eta=1$, $\eta=0.990$ and $\eta=0.936$.
Each plot shows the dependence on the phase $\varphi$ and the amplitude ratio $\tan \alpha$.
The highlighted colored lines represent the parameter values that are set in the various measurements, see Fig.~\ref{fig::dataUniversality} and Fig.~\ref{fig::eta_dependence_plot} below.
}
\label{fig::wv_parameters}
\end{figure*}

From Eq.~[\ref{eq::preSelEff}], we see that the overlap $\eta$ quantifies the loss of coherence between the two arms of the interferometer due to interactions and imperfections, which consequently leads to a reduction of the maximally achievable weak value.
The dependence of the weak value [\ref{eq::wvModified}] on $\eta$ as well as on $\alpha$ and $\varphi$ is presented in Fig.~\ref{fig::wv_parameters}.
Figs.~\ref{fig::wv_parameters}a,b show the case with ideal overlap $\eta = 1$, while Figs.~\ref{fig::wv_parameters}c-f illustrate the dependence for the non ideal case with reduced overlap and thus smaller $\left( {\rm \bf P}_A \right)_w$.

The weak value [\ref{eq::wvModified}] which accounts for multiple and even strong interactions is not useful to describe the whole of the external systems when inserted into expansion [\ref{eq::iaAmp}] because $\epsilon$ is large.
However, Eq.~[\ref{eq::wvModified}] can be used to describe the modification for those interactions which are weak, even if some of the other interactions or all of them together are arbitrarily strong.
We will show this now.

In our scenario we neglect the interactions of external systems in arm $A$ among themselves.
If between some particular  systems the interaction cannot be neglected, they are considered as a single composite system.
Thus, the interactions [\ref{eq::iaSingle}] in a single channel (Fig.~\ref{fig::MZI_basic}a) can be decomposed as
\begin{align}\label{eq::productPointers}
|\chi\rangle = \bigotimes_j  |\chi_j\rangle \rightarrow |\chi^\prime\rangle = \bigotimes_j  \eta_j \left(| \chi_j \rangle + \epsilon_j |\chi^\perp_j \rangle \right).
\end{align}
Here, as for [\ref{eq::iaSingle}], we absorbed the phases in the definitions of states, such that $\epsilon_j$ and $\eta_j$ are positive numbers.

In the case where the coupling to the, say, $k$-th system is weak, the change of the state of the system can be also expressed using density matrix language in the $\left\lbrace |\chi_k\rangle, |\chi^\perp_k\rangle \right\rbrace$ basis as
\begin{align} \label{eq::matOrig}
\rho_k =\begin{pmatrix}
1 & 0 \\
0 & 0
\end{pmatrix}  \rightarrow \rho^\prime_k = \begin{pmatrix}
1 & \epsilon_k \\
\epsilon_k & 0
\end{pmatrix} + \mathcal{O}(\epsilon^2_k).
\end{align}
For a particle passing through the MZI, when both the pre- as well as the postselection state are superpositions of $|A\rangle$ and $|B\rangle$, several interactions in $A$ (Fig.~\ref{fig::MZI_basic}b) will lead to entanglement between the various external systems.
Thus, each of the systems will be described by a mixed state.
The modified evolution of the weakly coupled $k$-th system is
\begin{align} \label{eq::matOrig1}
\rho_k \rightarrow \tilde{\rho}_k =  \begin{pmatrix}
1 & \left( {\rm \bf P}_A \right)^\ast_w \epsilon_k \\
\left( {\rm \bf P}_A \right)_w \epsilon_k & 0
\end{pmatrix} + \mathcal{O}(\epsilon^2_k).
\end{align}
Again, the modification of the effect of the weak interaction is characterized by the weak value $\left( {\rm \bf P}_A \right)_w$.

\section{Manifestation of the trace as shifts in pointer states}
In the previous sections we described the trace a particle leaves as the appearance of an orthogonal component in the quantum state of external systems.
Another language, frequently closer to experimental evidence is the change in the expectation values of the external systems.
Given the small change due to interactions in Fig.~\ref{fig::MZI_basic}a, expressed in [\ref{eq::iaAmp1}], every observable $O$ of the external system changes its expectation value as
\begin{align}\label{eq::shiftOsingle}
\delta \langle O \rangle \equiv \langle \chi^\prime | O | \chi^\prime \rangle - \langle \chi | O | \chi \rangle = 2 \epsilon \,\mathrm{Re} \left[ \langle \chi |O| \chi^\perp \rangle \right] + \mathcal{O}(\epsilon^2).
\end{align}
Then, using [\ref{eq::iaAmp}] (or [\ref{eq::matOrig1}] respectively) we see that for the pre- and postselected particle (Fig.~\ref{fig::MZI_basic}b) the change in the expectation value of $O$ is modified according to
\begin{align}\label{eq::shiftOunivers}
\tilde{\delta} \langle O \rangle = 2 \epsilon \,\mathrm{Re} \left[ \langle \chi |O| \chi^\perp \rangle \left( {\rm \bf P}_A \right)_w \right] + \mathcal{O}(\epsilon^2).
\end{align}
This formula is universal - it is valid for every system which was coupled weakly in arm $A$ to the particle passing through the interferometer.

Eq. [\ref{eq::shiftOunivers}] represents a new result in a very general scenario.
Let us now focus on the less general but very common measurement situation, which is usually considered when treating weak values \cite{AV90}.
There, a single observable $O$ is the pointer variable $Q$, the pointer wavefunction $\chi (Q)$ is real and the interaction with the particle in the channel shifts the wave function in the pointer variable representation as
\begin{align}
\chi(Q)\rightarrow \chi^\prime(Q) = \chi(Q-\delta Q). \label{eq::measTypeWvFct}
\end{align}
Obviously, this also shifts the expectation value
\begin{align}
\delta \langle Q \rangle = \delta Q. \label{eq::measTypeExpVal}
\end{align}
In this scenario $\chi^\perp (Q)$ is also real, as well as $\langle \chi |Q| \chi^\perp \rangle$.
Then, a positive weak value $\left( {\rm \bf P}_A \right)_w$ just tells us how the effect of the interaction is amplified or reduced according to
\begin{align}\label{delQ}
\tilde{\delta} \langle Q \rangle \approx \delta Q \, \mathrm{Re} [\left( {\rm \bf P}_A \right)_w].
\end{align}
If $\left( {\rm \bf P}_A \right)_w$ is negative, it tells us that the pointer will be shifted in the opposite direction.

If the weak value is imaginary, the expectation value of the pointer position will not be changed.
However, an orthogonal component in the quantum state of the pointer will still appear.
It will manifest itself in the shift of the expectation value of the momentum $P_Q$ conjugate to $Q$
\begin{align}\label{delP}
\tilde{\delta} \langle P_Q \rangle \approx 2 \delta  Q  \ (\Delta P_Q)^2 \; \mathrm{Im}[\left( {\rm \bf P}_A \right)_w],
\end{align}
where $(\Delta P_Q)^2=\langle \chi | P_Q^2 | \chi \rangle - \langle \chi | P_Q | \chi \rangle^2$ and $\hbar=1$.

\begin{figure*}
\includegraphics[width=1\textwidth]{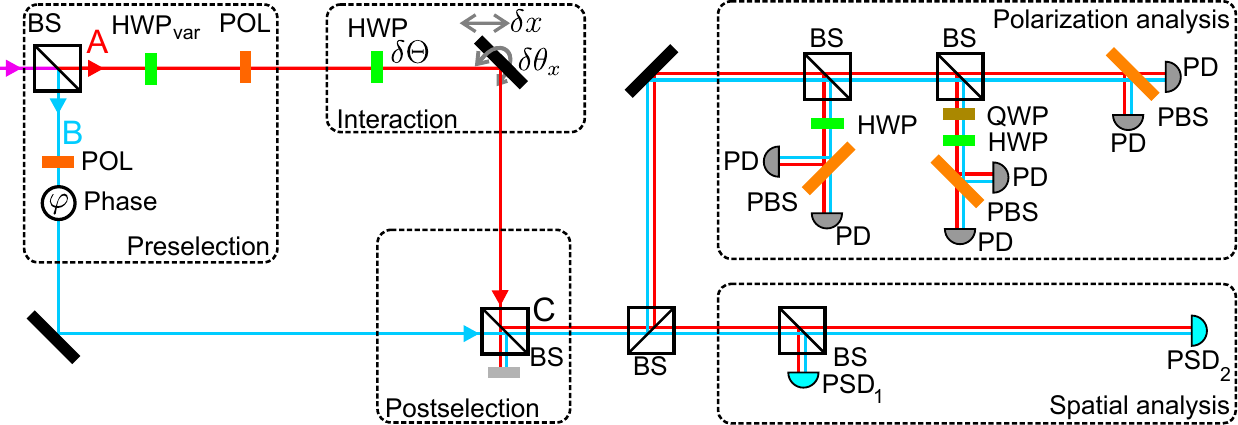}
\caption{
\textbf{Schematic experimental setup.}
The preselection state $| \psi \rangle$ is set using a non-polarizing beam splitter (BS) creating a spatial superposition between arms $A$ and $B$.
Two equally oriented polarizers (POL) and a half wave plate ($\mathrm{HWP}_{\mathrm{var}}$) are used to define the relative amplitudes.
Angle and position shifts, e.g. $\delta \theta_x$ and $\delta x$, are introduced by moving and tilting of optical components, whereas polarization rotations are imposed using a half wave plate (HWP).
The postselection is done by considering only one of the output ports ($C$) of the interferometer.
Analysis of the polarization degree of freedom is achieved by means of half and quarter wave plates (HWP and QWP), polarizing beam splitters (PBS), and photodiodes (PD), allowing the projection onto the polarization states $1/\sqrt{2}\left(|H\rangle \pm |V\rangle\right)$, $1/\sqrt{2}\left(|H\rangle \pm i |V\rangle\right)$, $|H\rangle$, and $|V\rangle$.
Position sensing detectors (PSD) at different $z$-positions allow to determine position and angle, respectively, in $x$ and in $y$ direction.
}
\label{fig::MZI_setup}
\end{figure*}

Eqs.~[\ref{delQ}] and [\ref{delP}] were obtained from [\ref{eq::shiftOsingle}] and [\ref{eq::shiftOunivers}] on the assumption of weak coupling when higher orders of $\epsilon$ can be neglected.
In the measurement situation [\ref{eq::measTypeWvFct}] with a Gaussian pointer, $\chi = e^{- Q^2/4(\Delta Q)^2}$ (we omit normalization), the usual range of validity of the weak value formalism is extended.
Even when the coupling is strong and the pointer distribution is significantly distorted during the measurement, the expressions for the shifts of the expectation values of $Q$, [\ref{eq::shiftOsingle}] and [\ref{eq::shiftOunivers}], remain exact, with
\begin{subequations}
\begin{align}
\delta \langle Q \rangle &= 2 \epsilon \mathrm{Re} \left[ \langle \chi |Q| \chi^\perp \rangle \right],  \label{eq::gaussianExactStd} \\
\tilde{\delta} \langle Q \rangle &= 2 \epsilon \mathrm{Re} \left[ \langle \chi |Q| \chi^\perp \rangle \left( {\rm \bf P}_A \right)_w \right]. \label{eq::gaussianExactWeakVal}
\end{align}
\end{subequations}
Indeed,  for the  Gaussian pointer $ \langle \chi |Q| \chi \rangle=0$, $ \langle \chi^\prime |Q| \chi^\prime \rangle=\delta Q$, and also the following expressions are easily calculated as
 \begin{equation}
\eta = \langle \chi |  \chi^\prime \rangle = e^{-(\delta Q)^2/8(\Delta Q)^2}, \quad \langle \chi |Q| \chi^\prime \rangle = \langle \chi |  \chi^\prime \rangle \frac{\delta Q}{2}.
\end{equation}
Then  [\ref{eq::gaussianExactStd}] is proven by substituting [\ref{eq::measTypeExpVal}] and [\ref{eq::iaSingle}], while including [\ref{eq::compPostSel}] and [\ref{eq::wvModified}] proves [\ref{eq::gaussianExactWeakVal}].

If the pointer is a Gaussian in the position variable $Q$ it is of course also a Gaussian in the conjugate momentum $P_Q$ representation.
Therefore, [\ref{delQ}] and [\ref{delP}], in analogy to the above, become exact formulas with $\Delta P_Q= \frac{1}{2 \Delta Q}$.
There are corresponding exact formulas for the effect of a shift in momentum $\delta P_Q$ with
\begin{subequations}
\begin{align}
\tilde{\delta} \langle P_Q \rangle  &= \delta P_Q \mathrm{Re} [\left( {\rm \bf P}_A \right)_w], \label{delPP} \\
\tilde{\delta} \langle Q \rangle &= -2 \delta P_Q  (\Delta Q)^2 \; \mathrm{Im}[\left( {\rm \bf P}_A \right)_w], \label{delQP}
\end{align}
\end{subequations}
see also \cite{Dressel2012}.

Direct substitution shows that the expressions remain correct for Gaussians in the regime of strong interactions also in the case of combinations of shifts in $Q$ and $P_Q$, such that
\begin{subequations}
\begin{align}
\tilde{\delta} \langle Q \rangle &= \delta  Q  \mathrm{Re} [\left( {\rm \bf P}_A \right)_w] - 2\delta P_Q  (\Delta Q)^2 \; \mathrm{Im}[\left( {\rm \bf P}_A \right)_w], \label{delQQ+P} \\
\tilde{\delta} \langle P_Q \rangle  &= \delta  P_Q \mathrm{Re} [\left( {\rm \bf P}_A \right)_w]  + \frac{\delta  Q}{2 (\Delta Q)^2} \; \mathrm{Im}[\left( {\rm \bf P}_A \right)_w]. \label{delPQ+P}
\end{align}
\end{subequations}
These equations are the basis of the alignment method presented in Section \ref{sec::alignment}.

\section{Observing the universality property} \label{sec::univ}
We use an optical Mach-Zehnder interferometer to experimentally visualize our central claim, namely, that all kinds of small effects of spatially pre- and postselected systems taking place at a specific location are modified in a universal manner characterized by the weak value of spatial projection.
In the experiment we demonstrate the universal change for three different couplings.
In every case the effect is modified in the same manner.

There are proposals and actual experiments where the photon couples to other particles in one arm of the interferometer \cite{Simon2011,Feizpour2011,Fu2015,Ben-Israel2017,Steinberg}.
In \cite{Steinberg} one arm of the interferometer is a Kerr medium and the photon passing through this arm changes the quantum state of the pointer by introducing a shift in the relative phase  between the wave packets of the pointer photons.
As it is done in most weak measurement experiments, instead of coupling to external particles we rather study interactions of the photons in an arm of the interferometer by observing the effect on other degrees of freedom of the photons itself.
We also used a (weak) laser beam, so all the results can be explained using Maxwell equations (although in a much more difficult way), but the observations would not change by employing single photons since intensity measurements are in one-to-one correspondence to single photon probability distributions.

The interactions in arm $A$ are realized by introducing controlled changes of spatial and polarization degrees of freedom.
The initial state of the position degree of freedom can be well approximated by a Gaussian along the $x$ as well as the $y$ coordinates.
The interaction is implemented by shifting the center of the Gaussian intensity distribution of the light beam going through arm $A$ by $\delta x$ compared to the beam going through arm $B$,
\begin{align}\label{spatialx}
\chi_x(x) =  e^{-x^2 / w^2_0} \rightarrow \chi^\prime_x(x) =  e^{-(x - \delta x)^2 / w^2_0},
\end{align}
where $w_0$ denotes the waist of the beam and normalization factors are omitted.

\begin{figure*}
\includegraphics[width=0.98\textwidth]{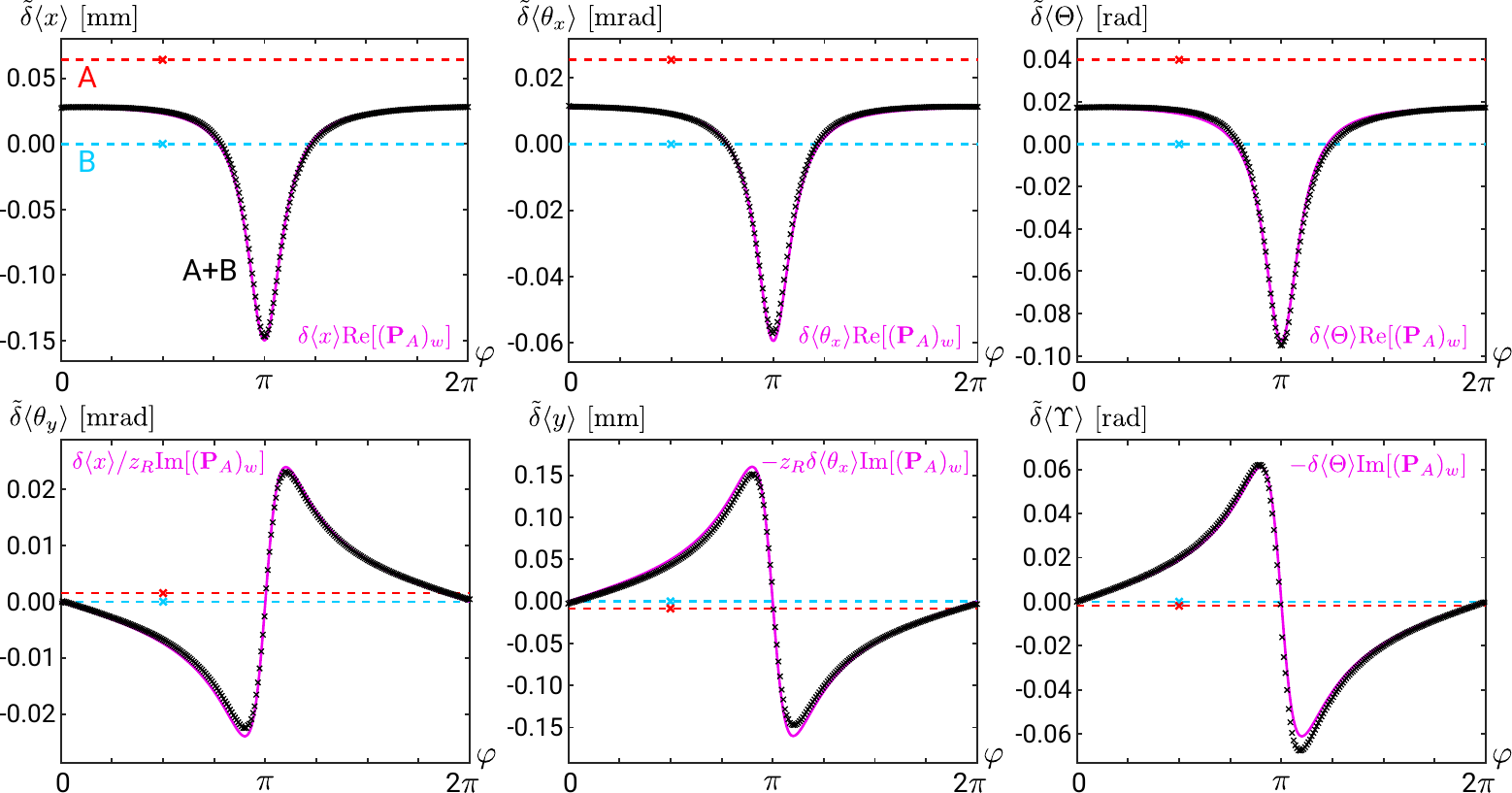}
\caption{\textbf{Observed Universality.}
(Upper row) The introduced displacements of arm $A$ in $x$ direction, angle around $x$-axis, and angle of polarization $\Theta$ ($\delta x$, $\delta \theta_x$, $\delta \Theta$) can be seen from the single red datapoints plotted at an arbitrary phase position.
The blue datapoints corresponding to arm $B$ are taken as a reference and thus show zero shift.
The axes are scaled such that the readings of $A$ agree for the three external systems.
For each of these three, the same behavior of the interference signal (black datapoints) is observed for the shifts of the variables $\tilde{\delta} \langle x \rangle$, $\tilde{\delta} \langle \theta_x \rangle$, and $\tilde{\delta} \langle\Theta\rangle$:
the effect seen from the measurement of the single arm is multiplied with the phase dependent real part of the weak value.
(Lower row) The analogous plots for the shift of the respective conjugate variables represented by $\tilde{\delta} \langle \theta_y \rangle$, $\tilde{\delta} \langle y \rangle$, and $\tilde{\delta} \langle\Upsilon\rangle$ show nicely the dependence on the imaginary part of the weak value.
The violet theoretical curves represent the rescaled real and imaginary parts of the weak value (no fit).
}
\label{fig::dataUniversality}
\end{figure*}

Another degree of freedom is the spatial state in $y$ direction of the light beam, which we modified by changing the angle of the beam around the $x$ axis, which for small angles corresponds to the momentum shift $\delta p_y = \frac{2\pi}{\lambda} \delta \theta_x$.
The resulting modification in arm $A$ can be expressed by
\begin{align}\label{spatialy}
\chi_y(p_y) =  e^{- w^2_0 p_y^2 / 4} \rightarrow \chi^\prime_y(p_y) = e^{- w^2_0 (p_y - \delta p_y)^2 / 4}.
\end{align}

As a third external system we use the photon polarization.
The interaction parameter here is the rotation of polarization by the angle $\delta \Theta$,
\begin{align}
|\chi_{\sigma} \rangle = | H \rangle \rightarrow |\chi_{\sigma}^\prime \rangle = \cos \frac{\delta \Theta}{2}  |H \rangle +\sin \frac{\delta \Theta}{2} |V \rangle,
\end{align}
where the states $| H \rangle$ and $| V \rangle$ are defined via $\sigma_z | H \rangle = | H \rangle$ and $\sigma_z | V \rangle = -| V \rangle$ for the Pauli matrix $\sigma_z$.

All other properties of the photon are expressed in the state $|\chi_{O} \rangle$.
Any imperfections of the interferometer can be understood to lead to a change of the initial state of these properties in arm $A$, $|\chi_{O} \rangle \rightarrow |\chi^\prime_{O} \rangle$ .

It is a good approximation to assume that there are no interactions between the external degrees of freedom we consider and thus we can express the quantum state of the photon in arm $B$ just before reaching the final beam splitter of the interferometer as
\begin{align}\label{eq::armB}
| B \rangle |\chi\rangle =  | B \rangle|\chi_x \rangle |\chi_{y} \rangle |\chi_{\sigma} \rangle |\chi_{O} \rangle,
\end{align}
while in arm $A$ it is
\begin{align}\label{eq::armA}
| A \rangle |\chi^\prime\rangle =  | A \rangle |\chi^\prime_x \rangle |\chi^\prime_{y} \rangle |\chi^\prime_{\sigma} \rangle |\chi^\prime_{O} \rangle.
\end{align}

To test the universality of modifications of effects for various degrees of freedom one could either perform complete tomographies of the final pointer states [\ref{eq::matOrig}] and [\ref{eq::matOrig1}] or, more clearly, show the modification of the effects of the three couplings according to [\ref{delQ}] and [\ref{delP}].
We follow the second approach.
More explicitly, we test the differences between effects of the interactions on the expectation values in three degrees of freedom  when the particle passes through the single arm (expressed by $\delta$) and when the particle passes through both arms (expressed by $\tilde \delta$)\footnote{We chose this method since our measurements of the shifts $\delta \langle x \rangle$, $\delta \langle \theta_x \rangle$, and $\delta \langle \Theta \rangle$ in a single channel are more precise than our control of the shifts $\delta x$, $\delta \theta_x $, and $\delta  \Theta $ via manual stages.}.
Because of the linear relation between $\theta_y$ and $p_x$ as well as $\theta_x$ and $p_y$, one obtains
\begin{eqnarray}
\tilde{\delta}\langle x \rangle =& \delta \langle x \rangle \mathrm{Re} [\left( {\rm \bf P}_A \right)_w], \label{delQx} \\
\tilde{\delta}\langle \theta_y \rangle  =& \frac{\delta \langle x \rangle}{z_R}\mathrm{Im}[\left( {\rm \bf P}_A \right)_w], \label{delPx} \\
\tilde{\delta}\langle \theta_x \rangle =& \delta \langle \theta_x  \rangle\mathrm{Re} [\left( {\rm \bf P}_A \right)_w], \label{delPy} \\
\tilde{\delta}\langle y \rangle =& -z_R \delta \langle \theta_x \rangle \mathrm{Im}[\left( {\rm \bf P}_A \right)_w]. \label{dely}
\end{eqnarray}
Here we have used the Rayleigh range $z_R \equiv \frac{\pi w^2_0}{\lambda}$ as the characteristic parameter of the Gaussian beam.

\begin{figure*}
\includegraphics[width=1\textwidth]{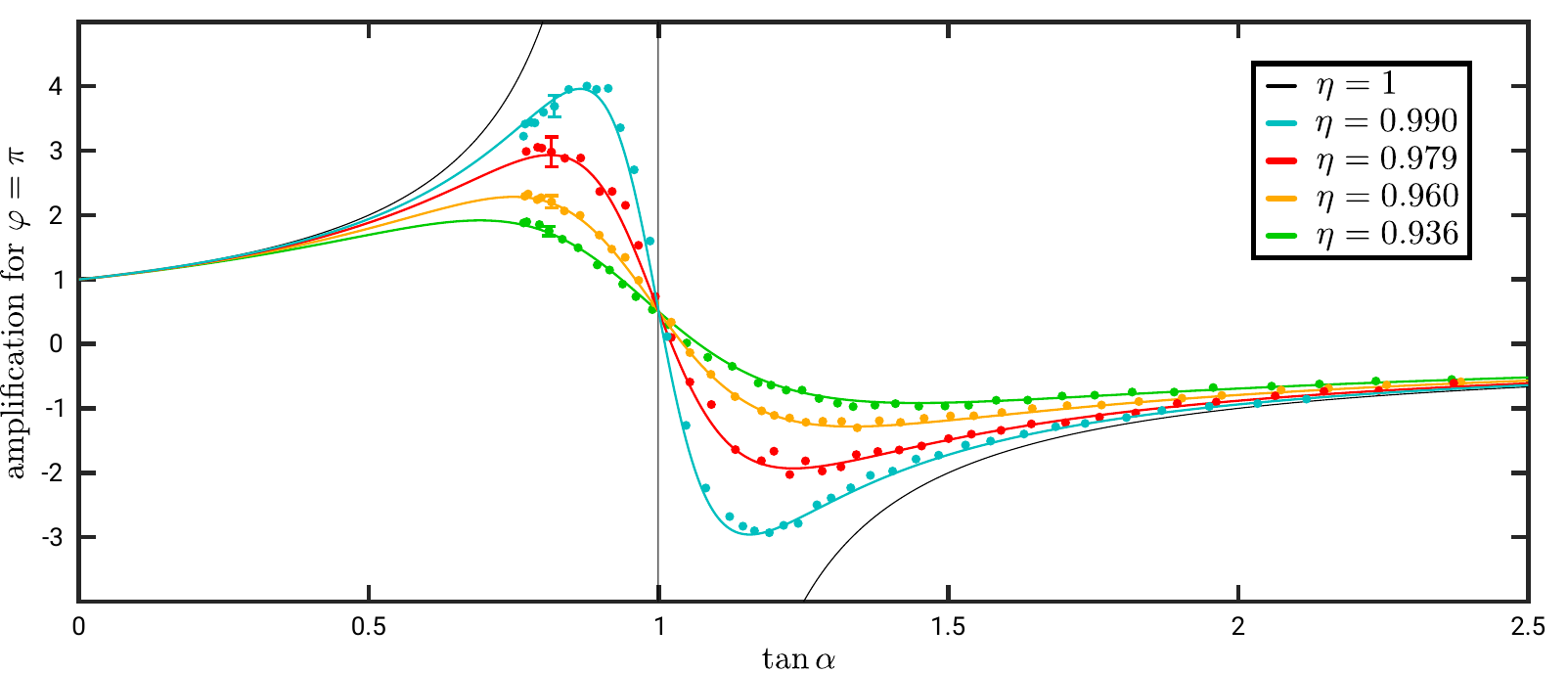}
\caption{\textbf{Modification of weak value due to decoherence.}
The colored dots represent the measured values for the modification of the shift $\delta x$ in the interference signal when varying the weak value via the relative amplitudes of the paths $A$ and $B$ ($\tan \alpha$ in Eq. [\ref{eq::wvModified}]) and fixed $\varphi = \pi$.
The four datasets correspond to four different values of the overlap $\eta$, which quantifies the coherence between the states of the external systems from the two arms.
The lines are theoretical curves as highlighted by the colored lines in Fig.~\ref{fig::wv_parameters}c,e.
Respective average error bars are shown for each $\eta$ on one of the first data points.
For comparison also the theoretical line with $\eta = 1$ (Fig.~\ref{fig::wv_parameters}a) is shown.
}
\label{fig::eta_dependence_plot}
\end{figure*}

The conjugate variable to the angle $\Theta$ defining polarization changes in the $\sigma_x$-$\sigma_z$ plane is an angle $\Upsilon$ describing polarization rotations in the $\sigma_y$-$\sigma_z$ plane relative to the initial state $| H \rangle$.
For small deviations these angles relate linearly to $\langle \sigma_x \rangle$ and $\langle \sigma_y \rangle$, respectively, and are given by
\begin{align}
\tilde{\delta} \langle \Theta \rangle &= \delta \langle \Theta \rangle  \mathrm{Re} [\left( {\rm \bf P}_A \right)_w], \label{delQTheta} \\
\tilde{\delta} \langle \Upsilon \rangle  &= - \delta \langle \Theta \rangle \mathrm{Im}[\left( {\rm \bf P}_A \right)_w]. \label{delPEpsilon}
\end{align}

The test was performed for the full range of $\varphi$ and thus for a large range of values $\left( {\rm \bf P}_A \right)_w$, see violet lines on the graphs of Fig.~\ref{fig::wv_parameters}.
The parameters for the calculation of $\left( {\rm \bf P}_A \right)_w$ necessary for testing relations [\ref{delQx}] - [\ref{delPEpsilon}] were also obtained from measurements. The signals  from separate arms (when the other arm was blocked) provided $\tan \alpha$.
The phase $\varphi$ and the overlap $\eta$ were obtained from the intensity of the interference signal and visibility measurements, respectively.
The relation between the visibility $\mathcal{V}$ and the overlap $\eta$ for the phase dependent output intensity $\mathcal{I} \propto \langle \phi | \rho |\phi \rangle \propto  1 + \tan^2 \alpha + 2 \tan \alpha \, \eta \cos \varphi$ is given by
\begin{align}
\mathcal{V} &\equiv \frac{\mathcal{I}_\text{max} - \mathcal{I}_\text{min}}{\mathcal{I}_\text{max} + \mathcal{I}_\text{min}} = \eta \frac{2 \tan \alpha}{1 + \tan^2 \alpha}.
\end{align}

The experiment is shown schematically in Fig.~\ref{fig::MZI_setup}.
After propagation through a single mode fiber for spatial filtering the horizontally polarized light from a laser diode ($\lambda=780\,\mathrm{nm}$) is split by a non-polarizing beam splitter.
The moduli of the amplitudes of the preselection state [\ref{eq::preSelSimp}] are controlled by means of rotating the polarization using a half wave plate in arm $A$ followed by a horizontal polarization filter.
The relative phase between the arms $\varphi$ is set by an optical trombone system with retroreflecting prisms moved by a piezoelectric crystal (not shown).

This setup enables to directly implement the three desired interactions along beam $A$ and simultaneuously measure their effect.
Fig.~\ref{fig::MZI_setup} depicts the setup.
The spatial displacement $\delta x$, which is schematically depicted as a shift of the mirror, was achieved by lateral movement of the prism from the trombone system.
Instead of a vertical tilt of this mirror, we incorporate the vertical rotation $\delta \theta_x$ by tilting the second beam splitter.
The polarization rotation $\delta \Theta$ is controlled by rotating a half wave plate in arm $A$.
Detecting light only from the output port $C$ provides the post-selection onto state $|\phi\rangle$, Eq.~[\ref{eq::postSelSimp}].

The photons at port $C$ are distributed onto several detectors using beam splitters for position and polarization analysis.
A position sensing detector $\mathrm{PSD}_1$ placed near the interferometer and a detector $\mathrm{PSD}_2$ placed farther away allows the estimation of position and angle in $x$ and $y$ directions.
We perform tomography of the polarization state using half and quarter wave plates in combination with polarizing beam splitters as shown in Fig.~\ref{fig::MZI_setup}.

A measurement run consists of three steps, namely, first a measurement of light propagating in arm $A$ alone, second of arm $B$ alone, and last a measurement of the interference signal.
The six expectation values obtained from measurements of arm $B$ are used as a reference for the subsequent analysis.

The measurement with only beam $A$ shows the effect of the interactions when the photons pass through a single channel as in Fig.~\ref{fig::MZI_basic}a.
The results are indicated in the graphs of Fig.~\ref{fig::dataUniversality} as red dashed horizontal lines since they exhibit no dependence on the phase\footnote{Please contact J.D. (jan.dziewior@physik.lmu.de) if you desire access to the raw experimental data for this plot as well as for all other plots.}.

The universality is clearly shown by the similarity of the results for the three couplings (Fig.~\ref{fig::dataUniversality}).
Of course in all graphs the observed values are different and have different units.
For demonstration purposes we arranged the scales of the graphs in the upper row of Fig.~\ref{fig::dataUniversality} such that the signals of all interactions, $\langle x \rangle_A$, $\langle \theta_y \rangle_A$, $\langle \Theta\rangle_A$ have the same size.
We were trying to avoid shifts in conjugate variables as much as possible.
Our measurement results, red dashed lines in the plots from the lower row of Fig.~\ref{fig::dataUniversality}, show that the tuning was good, although not perfect.

\begin{figure*}
\includegraphics[width=1\textwidth]{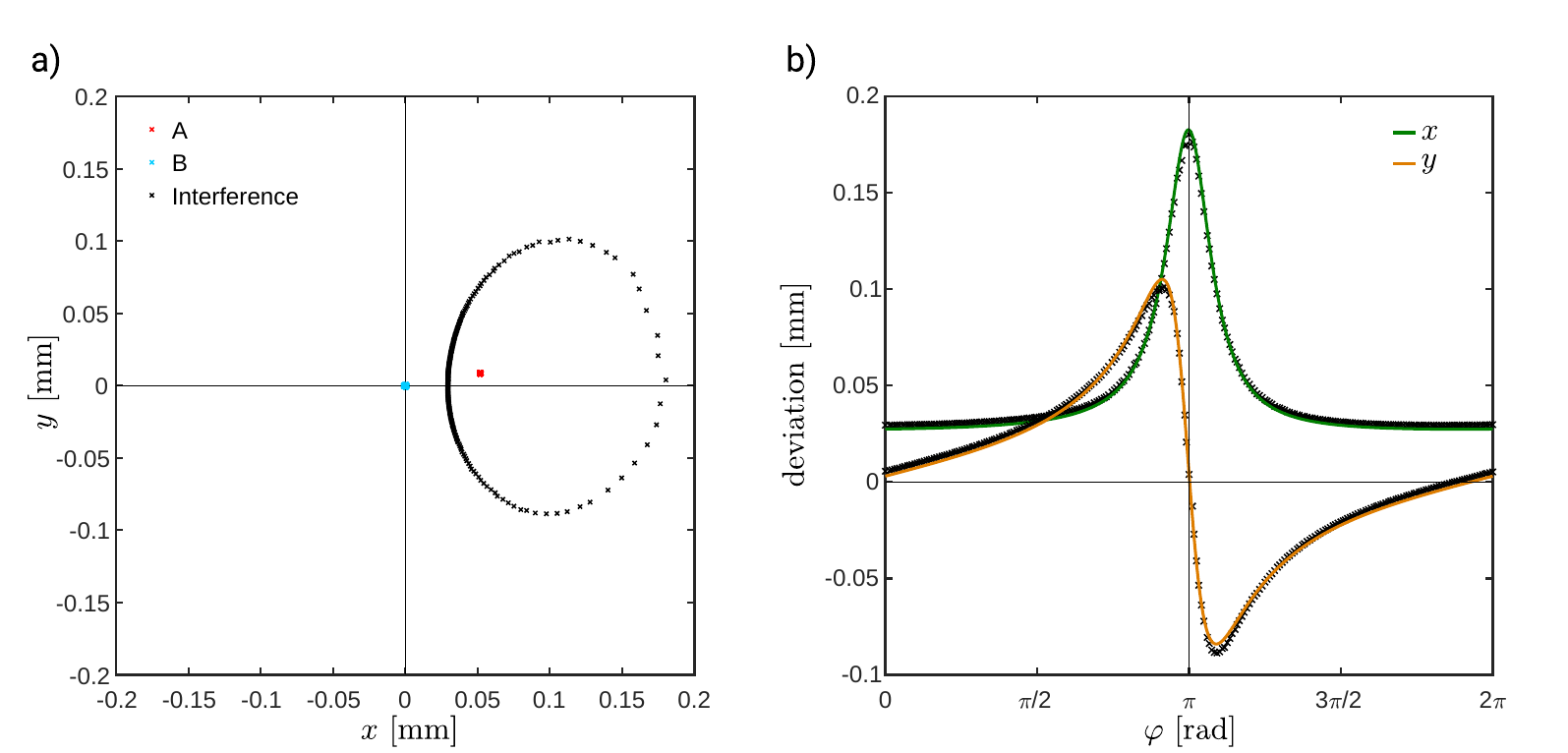}
\caption{\textbf{a) Trajectories of beam centroids in output $C$ for misaligned MZI.}
The \textit{blue} and \textit{red} spots correspond to the measurements of the beams from the single arms when the other arm is blocked.
While the \textit{blue} spot at the origin corresponds to beam $B$ without interaction, the \textit{red} spot corresponds to the misaligned beam $A$.
The elliptic trajectory of the interference pattern is represented by the \textit{black} points.
\textbf{b) Fits onto $x$ and $y$ projections of trajectory.}
By fitting the  vector function [\ref{eq::alignFormulaCoarse}] to the $x$- and $y$-projections of the interference ellipse we determine the parameters of the misalignment.
}
\label{fig::trajAlignment}
\end{figure*}

\begin{figure}
\includegraphics[width=0.5\textwidth, height=0.45\textwidth]{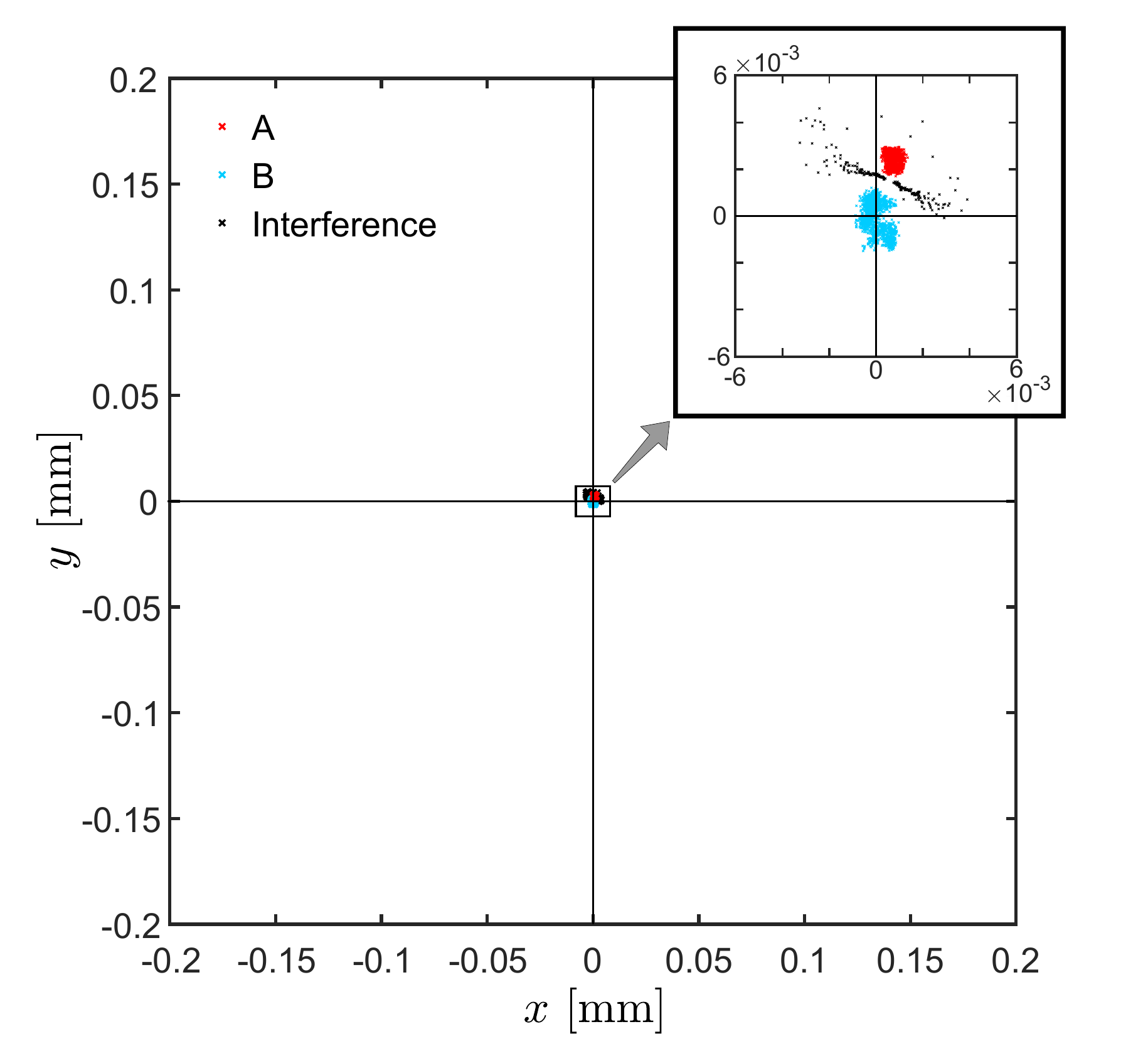}
\caption{\textbf{Trajectories of beam centroids after one alignment step.}
It can be clearly seen how size of the ellipse and the distance between the centroids of the single beams $A$ (\textit{red}) and $B$ (\textit{blue}) are significantly reduced in comparison to Fig.~\ref{fig::trajAlignment}a.
}
\label{fig::trajAfter}
\end{figure}

Continuous violet lines on these graphs provide theoretical predictions based on the weak value  $ \left( {\rm \bf P}_A \right)_w $ given by [\ref{eq::wvModified}] and the interactions in the single arms are presented as red and blue dashed lines in the graphs.
The intensities obtained measuring arm $A$ and arm $B$ alone yield $\tan\alpha = 1.3323 \pm 0.0002$.
From the visibility measurement, ${\cal V} = 95.09 \pm 0.02\%$, we obtained $\eta=0.9904 \pm 0.0003$.
For these parameters we observed amplifications with factors up to $4$ and $-3$.
The very good agreement between experimental data and theoretical predictions, shown in Fig.~\ref{fig::dataUniversality}, demonstrates the universality of the modification of several fundamentally distinct forms of interactions for couplings with a pre- and postselected system.

To evaluate the dependence of the weak value on the coherence between the two arms parametrized by $\eta$, we measured the effect of the displacement in $x$ on the output beam.
For this run we kept the phase fixed at $\varphi = \pi$ and varied the amplitude ratio $\tan \alpha$ covering another region of the parameter space from Fig.~\ref{fig::wv_parameters}.
We changed the coherence by varying the polarization misalignment leading to a smaller overlap between the photon states passing through the two arms.
The modification of the shift in $x$ direction presented in Fig.~\ref{fig::eta_dependence_plot} follows nicely the weak value [\ref{eq::wvModified}].

\section{Alignment Method} \label{sec::alignment}
In the previous sections we considered a scenario in which the path state of a photon in an arm of an interferometer is coupled to its other degrees of freedom, in particular its spatial degrees of freedom in $x$ and $y$ direction.
This scenario exactly represents a situation encountered in real experimental interferometric setups, namely when the arms of the interferometer are misaligned.
The differences in position $\delta \vec{r} \equiv \left( \delta x, \; \delta y \right)$ and angle $\vec{\delta \theta} \equiv \left( \delta \theta_x, \; \delta \theta_y \right)$ between the photons passing through distinct arms of the interferometer can be considered as results of interactions in one arm, which change the initially identical spatial states of the particle.

It is well known that the picture generated by the interference of the beams from a misaligned interferometer displays a strong phase dependence.
Fig.~\ref{fig::trajAlignment}a shows the centroid trajectory during the phase scan of a misaligned interferometer.
We demonstrate that it is possible to quantitatively determine the exact misalignment parameters of the interferometer by analyzing this phase dependent movement.
In fact, the misalignment parameters $\delta \vec{r}$ and $\vec{\delta \theta}$ could be calculated from measurements described in the previous section.
Disregarding the polarization analysis it was a measurement of the misalignment parameters based on position measurements of centroids of the beams on two detectors at different locations.
But the method is more powerful and can be implemented with only a single position sensitive detector as well.

The basis for our alignment method are Eqs.~[\ref{delQQ+P}] and [\ref{delPQ+P}] which, somewhat surprisingly, remain precise even for large misalignment.
The shift observed on the single detector $\tilde \delta \vec{R}$ is the sum of the position shift $\tilde \delta \vec{r}$ and the position shift due to the shift in direction $\vec{\delta \theta}\times\vec{L}$, where $\vec{L} = (0,0,z)$ is the vector parallel to the beam with the length equal to the distance $z$ along the beam between the waist and the detector.
Thus, the position shift of the centroid on the detector $\tilde{\delta} \vec{R}$ is given by
\begin{align}\label{eq::alignFormulaCoarse}\nonumber
\tilde \delta\vec{R} &= \left(\delta x + z \delta \theta_y,~\delta y - z \delta \theta_x \right) \mathrm{Re}[\left( {\rm \bf P}_A \right)_w] + \\
&\left(\frac{z}{z_R}\delta x - z_R \delta \theta_y,~\frac{z}{z_R} \delta y + z_R \delta \theta_x \right) \mathrm{Im}[\left( {\rm \bf P}_A \right)_w].
\end{align}
The weak value is given by [\ref{eq::wvModified}].
The parameters $\tan \alpha$, $\eta$, $z$, and $z_R$ are found experimentally as in the previous section.
The function [\ref{eq::alignFormulaCoarse}] corresponds to the trajectory of the beam centroid on the detector as shown in Fig.~\ref{fig::trajAlignment}a.
Even small misalignments which otherwise might be difficult to resolve become detectable due to the effect of weak amplification.

Fig.~\ref{fig::trajAlignment}b shows the $x$- and $y$-components of $\tilde \delta\vec{R}$ as functions of $\varphi$.
A least squares fit of this function provides the four unknown misalignment parameters $\delta \vec{r}$ and $\vec{\delta \theta}$.
It is remarkable that a fit function with so few parameters accurately models the experimental results.
For the data shown the fit provided $\delta \vec{r} = (49 \pm 2,7 \pm 2) \,\mathrm{{\mu}m}$ and $\vec{\delta \theta} = (12.7 \pm 0.4,0.2 \pm 0.4)\,\mathrm{{\mu}rad}$.

We have performed corrections according to these parameters and repeated our procedure, see Fig.~\ref{fig::trajAfter}.
The stability of the centroid shows excellent alignment and a subsequent fit procedure provides the parameters $\delta \vec{r} = (-1 \pm 2,2 \pm 2) \,\mathrm{{\mu}m}$ and $\vec{\delta \theta} = (0.2 \pm 0.4,-0.6 \pm 0.4)\,\mathrm{{\mu}rad}$.

In our method to obtain the misalignment parameters we rely on the knowledge of the beam parameters, i.e., Rayleigh range $z_R$ and longitudinal position of the detector relative to the waist $z$.
In some situations the reversed task might be of interest.
If we control the misalignment parameters, we can also use our algorithm with the fit to obtain the beam parameters.

In fact, the general idea of alignment using weak values was already used in alignment of the interferometer demonstrating the past of a particle in nested interferometers \cite{Danan} and since then it was significantly developed and improved \cite{DimaMSc,NimrodMSc} until it reached the efficiency presented in the current work when a single scan led to a very good alignment.

\section{Trace and Presence}\label{sec::discussion}
A generic property of weak measurements is the possibility to perform several weak measurements on the same system.
Thus, we can interpret our experiment as multiple weak measurements of the projection operator which all yield the same result, the weak value of the
projection on the arm of the interferometer.
However, it also implies a broader meaning with respect to the discussion of the local presence of quantum particles.

A classical particle can either be in a particular location or not.
The presence of a quantum particle in a certain location, however, is a subtle issue and its analysis strongly depends on the adopted interpretation of quantum mechanics.
To avoid controversial interpretational issues, we do not discuss ontological aspects of the concept of presence of a particle and instead argue within the operational approach.

When the wavefunction of a quantum particle is well localized in a particular location, the trace is specified in a unique way by the local interaction in that location in analogy to the trace of a classical particle when it is present, see Eq.~[\ref{eq::iaSingle}].
Given that there are only local interactions in nature, there is no trace when the wavefunction vanishes.
Similarly, there is no trace in a classical channel when the particle  is absent.

Scenarios when the wavefunction does not vanish, but is also not fully localized at this location, are no longer understandable from a classical perspective.
The universal relation between the trace in these scenarios and the trace of a fully localized particle which we found in our work can be considered
as a basis of an operational concept of presence of a particle.
It goes beyond defining the particle as present when it leaves a trace and not present, when it does not \cite{past}.

According to our operational approach, the ``presence'' of a pre- and postselected particle in the arm $A$ of an interferometer is defined according to the way it affected the external systems to which it was coupled and is quantified by the complex number $(\mathbf{P}_A)_w$.
This definition yields ``presence'' 1 when the forward evolving wavefunction of the particle is solely inside the arm $A$ independently of postselection, but ``presence'' 1 can happen also when neither the pre- nor the postselected states are eigenstates of the local projection on arm $A$.
The ``presence'' 0 or ``absence'' of the particle is ensured when the forward evolving wavefunction of the particle vanishes in arm $A$, but it is not a necessary condition.
The postselected particle might have been ``absent'' in arm $A$ (no effect on local external systems can be observed) even when the forward evolving wavefunction did not vanish there.

Our concept provides a quantification and characterization of presence by describing the modification of effects of the particle's interactions with external systems.
It can be increased, decreased, or changed in a particular, well defined way and this change is the same for all local interactions - it is universal.

\section{Conclusions}
We have analyzed theoretically and experimentally the modifications of the effect of weak interactions on pre- and postselected particles.
We have shown that there is a universal description of the modification of these couplings for all weak interactions given by a single complex number, the weak value of the projection on the corresponding location.

Our approach is based on expressing the effect of external systems in terms of the orthogonal components which appear due to the interactions.
This allows to formalize the meaning of the weak value without reference to a specific form of coupling.
The weak value not only modifies a shift of expectation values as usual, but also the relative amplitudes of the orthogonal components of all external quantum systems interacting with the particle.

The experiment shows for three different couplings that each of the effects is modified in exactly the same way.
This is shown for not just a few cases of pre- and postselected particles, but for a continuum of parameters with a large range of weak values of projection.

The approach derives the general expression [\ref{eq::wvModified}] which allows to apply the concept of weak values for several couplings which are not necessarily weak.
These findings enable one to understand seemingly complicated dependencies seen in experiments, for example \cite{Kofman2012}, and can facilitate multi-parameter precision measurements in the future.

We define an operational paradigm for the presence of a pre- and postselected particle according to the trace it leaves.
It is more intricate than the dichotomic concept of the presence of a classical particle which can only be present or not.
This complexity is surprising in light of the fact that in all scenarios the external systems are in a superposition or a mixture of the undisturbed state with a single particular orthogonal component.

Our demonstration of the universality of the modification of the interactions led us to a novel alignment method.
Its effectiveness relies on the unexpected robustness of the modification of Gaussian pointers, where the weak value expressions remain precise even for strong couplings.
In our method a single phase scan suffices to recover all misalignment parameters from the analysis of the position of the centroid of a single output beam, clearly reducing the effort in an often tedious task, while at the same time potentially harnessing the benefits of weak value amplification.

\acknow{This work has been supported in part by the Israel Science Foundation Grant No. 1311/14, the German-Israeli Foundation for Scientific Research and Development Grant No. I-1275-303.14, the DFG Beethoven 2 Project No. WE2541/7-1, and by the German excellence initiative Nanosystems Initiative Munich. J.D. acknowledges support by the international Max-Planck-Research school for Quantum Science and Technology (IMPRS-QST), L.K. acknowledges support by the international Ph.D. program ExQM from the Elite Network of Bavaria, and J.M. acknowledges support of the LMU research fellowship.}

\showacknow{}

\bibliography{bib_universality}

\end{document}